\documentclass[conference]{IEEEtran}
\usepackage{cite}
\usepackage{flushend}
\usepackage{amsmath,amssymb,amsfonts}
\usepackage{algorithmic}
\usepackage{graphicx}
\usepackage{textcomp}
\usepackage{xcolor}
\def\BibTeX{{\rm B\kern-.05em{\sc i\kern-.025em b}\kern-.08em
    T\kern-.1667em\lower.7ex\hbox{E}\kern-.125emX}}
\begin{document}

\title{CT-To-MR Conditional Generative Adversarial Networks for Ischemic Stroke Lesion Segmentation}

\author{\IEEEauthorblockN{Jonathan Rubin}
\IEEEauthorblockA{\textit{Acute Care Solutions} \\
\textit{Philips Research North America}\\
Cambridge, MA, USA \\
jonathan.rubin@philips.com}
\and
\IEEEauthorblockN{S. Mazdak Abulnaga}
\IEEEauthorblockA{\textit{Computer Science and Artificial Intelligence Lab} \\
\textit{MIT}\\
Cambridge, MA, USA \\
abulnaga@mit.edu}}

\maketitle

\begin{abstract}
Infarcted brain tissue resulting from acute stroke readily shows up as hyperintense regions within diffusion-weighted magnetic resonance imaging (DWI). It has also been proposed that computed tomography perfusion (CTP) could alternatively be used to triage stroke patients, given improvements in speed and availability, as well as reduced cost. However, CTP has a lower signal to noise ratio compared to MR. In this work, we investigate whether a conditional mapping can be learned by a generative adversarial network to map CTP inputs to \emph{generated} MR DWI that more clearly delineates hyperintense regions due to ischemic stroke. We detail the architectures of the generator and discriminator and describe the training process used to perform image-to-image translation from multi-modal CT perfusion maps to diffusion weighted MR outputs. We evaluate the results both qualitatively by visual comparison of generated MR to ground truth, as well as quantitatively by training fully convolutional neural networks that make use of generated MR data inputs to perform ischemic stroke lesion segmentation. Segmentation networks trained using generated CT-to-MR inputs result in at least some improvement on all metrics used for evaluation, compared with networks that only use CT perfusion input.
\end{abstract}


\section{Introduction}
Ischemic stroke is caused by partial or total restriction of blood supply to part of the brain. During an acute stroke, prolonged ischemia, or insufficient blood supply, results in irreversible tissue death. Decisions about ischemic stroke therapy are highly time-sensitive and rely on distinguishing between the infarcted core tissue and hypoperfused lesions. As such, automated methods that can locate and segment ischemic stroke lesions can aid clinical decisions about acute stroke treatment. Computed tomography perfusion (CTP) has been used to triage stroke patients and has advantages in cost, speed and availability over diffusion-weighted magnetic resonance imaging (DWI). CTP provides detailed information about blood flow within the brain and can determine areas that are (in)adequately perfused with blood. However, CTP has a lower signal to noise ratio compared to DWI where infarcted core brain tissue readily shows up as hyperintense regions \cite{louvbld1997ischemic,van1994water}.

In this work we train generative adversarial networks to learn a conditional mapping that maps CTP infarcted core regions to more clearly delineated hyperintense areas in \emph{generated} MR scans. We aim to improve lesion segmentation results by augmenting CTP data with synthesized DWI images. We utilize a dataset of 94 paired CT and MR scans~\cite{cereda2016benchmarking} made available as part of the ISLES 2018 Ischemic Stroke Lesion Segmentation Challenge~\cite{maier2017isles,winzeck2018isles}. Data was collected from 63 subjects from 4 hospital sites worldwide. Each acute stroke patient underwent back-to-back CTP and MRI DWI imaging within 3 hours of each other, and DWI scans were used as ground truth for segmenting lesions. Each CT scan was co-registered with its corresponding DWI using Montreal Neurological Institute (MNI) tools \cite{cereda2016benchmarking}. Scans were acquired with a resolution of $1\times 1\times 5$ mm, where the axial plane was sampled sparsely. Perfusion maps were derived from each CT scan including cerebral blood flow (CBF), cerebral blood volume (CBV), mean transit time (MTT) and time to peak of the residue function (Tmax).

We employ the image-to-image translation framework introduced in~\cite{isola2016image} and modify it to accept multi-modal CT perfusion maps as input. After training conditional generative adversarial networks (CGANs) to reconstruct MR from CT perfusion maps, we train fully convolutional neural networks (FCN) to performs semantic segmentation of infarcted core tissue and compare whether performance can be improved by including the \emph{generated} MR as an extra channel of information to the network. We show that FCNs trained with combined CTP and generated MR inputs lead to overall performance improvement on a range of metrics, compared to networks trained without extra derived MR information.

\section{Related Work}
The high cost of acquiring ground truth labels means that many medical imaging datasets are either not large enough or exhibit large class imbalances between healthy and pathological cases. Generative models and GANs attempt to circumvent this problem by generating synthesized data that can be used to augment datasets during model training. Consequently, GANs are increasingly being utilized within medical image synthesis and analysis~\cite{wolterink2017deep,nie2017medical,lau2018scargan,wolterink2018blood}. We make use of the pix2pix framework~\cite{NIPS2014_5423} as it is able to learn both the image translation map and an appropriate loss function for training. The framework has been used effectively in several tasks, such as to simulate myocardial scar tissue in cardiovascular MR scans~\cite{lau2018scargan}. Here, GANs were used to augment healthy MR scans with realistic-looking scar tissue. Using two GANs, \cite{lau2018scargan} were able to both generate the scar tissue and refine intensity values using a domain-specific heuristic. The generation of scar tissue was effective in training and led to realistic results -- experienced physicians mistook the generated images as being real.

 \cite{wolterink2017deep} investigates the problem of CT generation from MR for radiation therapy planning -- a task that requires both MR and CT volumes. They synthesize CT from MR scans using CycleGAN~\cite{CycleGAN2017}, which employs a forward and backward cycle-consistency loss. The CycleGAN framework allows processing of unpaired MR and CT slices and does not rely on having co-registered images. Interestingly,~\cite{wolterink2017deep} shows that a model trained using unpaired MR and CT data was able to outperform models that used paired data.

3D CGANs were also used in \cite{jin2018ct} to learn shape information about pulmonary nodules in CT volumes. They generated synthetic nodules by training a generator on pairs of nodule-masked 3D input patches together with their ground truth segmentations. They used the generator network to augment their CT training set with nodules close to lung borders to improve segmentation. As in our work, and others ~\cite{isola2016image,lau2018scargan}, the authors of ~\cite{jin2018ct} use a combination of the L1 and conditional GAN loss to train their networks.

In addition to the works mentioned above, adversarial networks have also been used for improving lung segmentation in chest radiographs \cite{dai2018scan}, correcting motion-related artifacts in cardiac MR images \cite{oksuz2018cardiac},
and in registering medical images \cite{mahapatra2018deformable}. The motivation for our work is to explore whether hyperintensitities in MR scans can be emulated from CTP inputs with CGANs to improve the performance of ischemic stroke lesion segmentation.

\section{Contributions}

The contributions of this work are as follows:

\begin{enumerate}
\item We develop and train a modified CGAN network to reconstruct diffusion-weighted MR images from CTP maps.
\item We combine the generated MR images with the CTP data to train fully convolutional neural networks to perform segmentation of ischemic core stroke lesions.
\item We demonstrate the learned conditional mapping from the CTP input results in more clearly delineated hyperintense regions in generated MR, resulting in improved segmentation of the infarcted core compared to prior models.
\end{enumerate}

\section{CT-To-MR Conditional GAN Architecture}

\begin{figure*}[htbp]
  \includegraphics[height=0.6\linewidth]{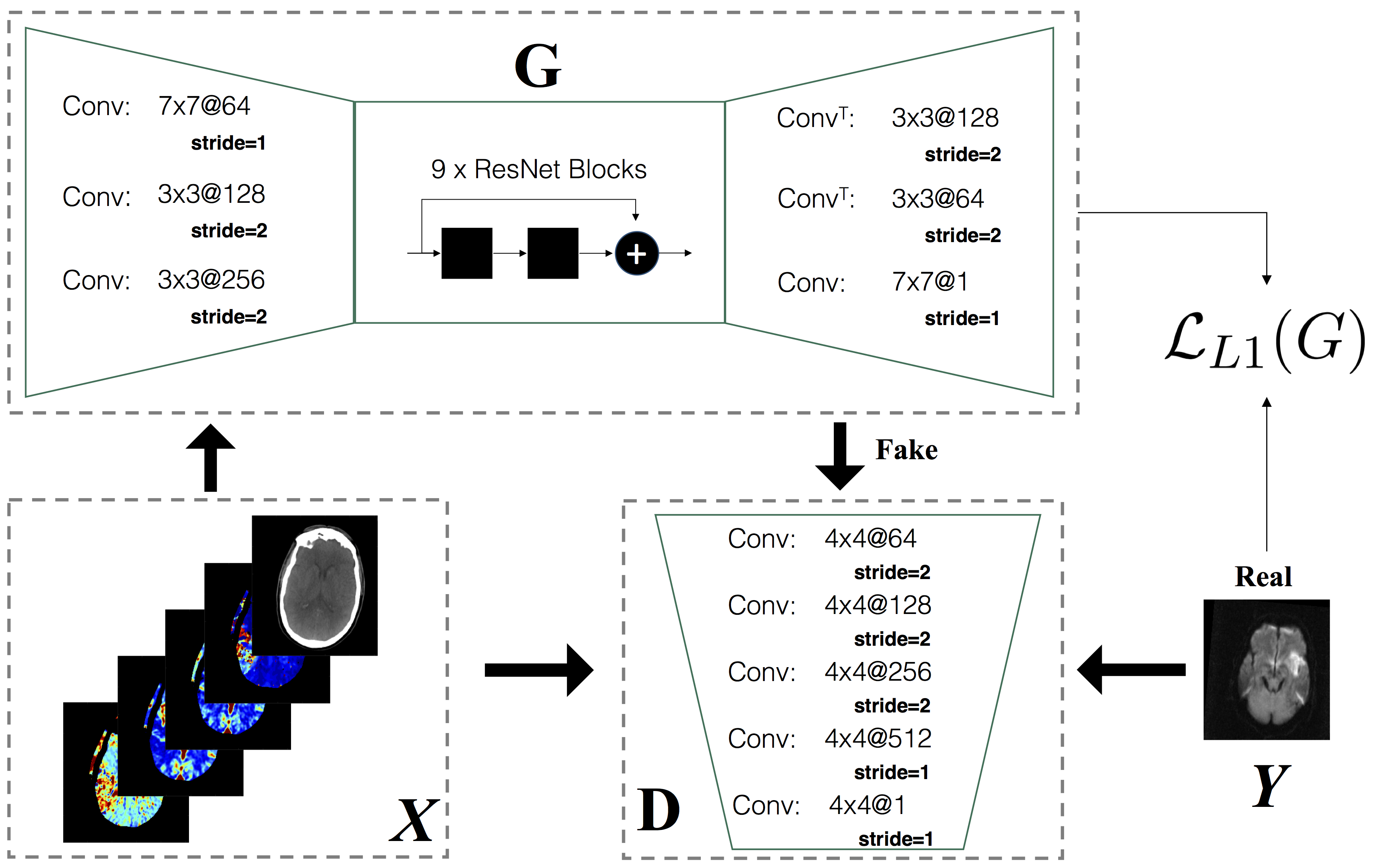}
  \caption{Overview of CT-To-MR Conditional GAN architecture.}
  \label{fig:arch}
\end{figure*}

Generative adversarial networks \cite{NIPS2014_5423} train both a generator and a discriminator network. The generator network, $G(z)$, attempts to generate outputs that resemble images, $y$, from a distribution of training data, where $z$ is a random noise vector, $G : z \rightarrow y$. The discriminator network, $D(\cdot)$, is given either a real input, $D(y)$, or a generated one, $D(G(z))$, and attempts to distinguish whether the input is real (images resulting from the true underlying data distribution) or fake (images created by the generator).

For the task of generating MR images conditioned on CT perfusion inputs, we adopt the Conditional GAN formulation, introduced in \cite{mirza2014conditional} and further described in the pix2pix framework \cite{isola2016image}. Conditional adversarial networks alter the generator such that it is conditioned on an original input image, $G: {x, z} \rightarrow y$, where $x$ is the input image and $z$ is once again a random noise vector. The discriminator function, $D(\cdot)$, is also updated to accept the conditional image, $x$, as input, as well as the real, $D(x, \bf{y})$, or fake input, $D(x, \bf{G(x, z)})$, created by the generator. The full objective function for the conditional generator is given in~\eqref{eqn:cgan}.

\begin{equation}
\begin{split}
\mathcal{L}_{CGAN}(G, D) = \mathbb{E}_{x, y\sim{p_{data(x,y)}}} [\log(D(x, y)] + \\
\mathbb{E}_{x\sim{p_{data(x)}, z\sim{p_z(z)}}}[\log(1 - D(x, G(x, z)))]
\end{split}
\label{eqn:cgan}
\end{equation}

The expected value in (\ref{eqn:cgan}) is taken over a batch of data and effectively results in the mean over a collection of  real data items, as well as fake data items. As in~\cite{isola2016image}, $z$ is introduced into the generator network in the form of dropout at both train and test time. The final objective function for training the CT-To-MR translation model combines both the global $\mathcal{L}_{CGAN}(G, D)$ loss together with an additional L1 loss term,~\eqref{eqn:l1}, that captures the local per-pixel reconstruction error. The combined objective function is given in~\eqref{eqn:objv}, where $\lambda$ is selected as a hyperparameter that provides a weighting over the separate loss terms.

\begin{equation}
\mathcal{L}_{L1}(G) = \mathbb{E}_{x, y\sim{p_{data(x,y)}}, z\sim{p_z(z)}} [\left \| y - G(x, z)\right \|_1]
\label{eqn:l1}
\end{equation}

\begin{equation}
G^* = \arg\min_{G}\max_{D} = \mathcal{L}_{CGAN}(G, D) + \lambda\mathcal{L}_{L1}(G)
\label{eqn:objv}
\end{equation}

\subsection{Generator Architecture}
A high level overview of the generator architecture, $G$, is shown in Fig.~\ref{fig:arch}. Generator inputs, $x$, are 5-channel $256\times256$ CT perfusion slices that contain the CT scan, stacked together with the CBF, CBV, MTT, and Tmax perfusion maps. We utilize 2D networks because the CTP scans were acquired only near regions with lesion, resulting in images with varying axial depth, ranging from $2-22$ slices. First, three initial convolution operations (Conv) are applied. The size and number of convolutional kernels are shown in the figure as: $n\times n$@$f$, where $n$ is the kernel size and $f$ the number of kernels. Downsampling is achieved via strided convolution. This is followed by 9 ResNet blocks, where a ResNet block is a residual block \cite{he2016deep} that consists of the following operations: Conv-InstanceNorm-ReLU-Dropout-Conv-InstanceNorm. In the preceding operations, instance normalization \cite{ulyanovinstance} is similar to batch normalization \cite{ioffe2015batch}, however, it results in an instance-specific normalization. Before each Conv operation in the block, reflection padding with size 1 is added to each border of the input. The number of feature maps stays constant at 256, throughout the 9 ResNet blocks, as does their spatial resolution. Upsampling is achieved in the generator via fractionally strided convolutions (Conv$^T$), as shown in Fig.~\ref{fig:arch}. The generator output is a $1\times256\times256$ single channel derived MR slice.

\subsection{Discriminator Architecture}

As in~\cite{isola2016image}, we utilize a convolutional PatchGAN discriminator that models high frequency image structure in local patches and penalizes incorrectness at the $N{\times}N$ patch-level, where $N$ is the size of the patch. This is combined with the $L1$ loss term, $\mathcal{L}_{L1}$, that penalizes low frequency distortion. A high level overview of the discriminator, $D$, is depicted in Fig.~\ref{fig:arch}.

The conditional discriminator accepts either real, $D(x, y)$, or generated, $D(x, G(x, z))$, MR slices, together with the original CT data and perfusion maps, $x \in \mathbb{R}^{5\times256\times256}$.
CTP data and `real' or `fake' DWI MR slices are stacked together in the channel dimension resulting in $6\times256\times256$ inputs being processed by the PatchGAN discriminator. All convolutions use a kernel size of $4\times4$, with downsampling once again being handled via strided convolution. Excluding the first and last convolution shown in Fig.~\ref{fig:arch}, each convolution is followed by an instance normalization operation and LeakyReLU activation with a negative slope coefficient of $0.2$. The output of the network is a $30\times30$ map of discriminator activations, where each activation captures a $70\times70$ receptive field of overlapping patches from input channels. The final discriminator output is given by an average of this activation map.

\subsection{CT-To-MR Conditional GAN Training}

\subsubsection{Data split}

From the 94 available scans, a 5-fold split of the dataset was performed to create $5 \times 80\%/20\%$ splits. Splits were performed \emph{by subject} to ensure that each fold of the data consisted of scans from unique subjects. To ensure the CGAN model generated ``MR slices'' only for scans it was not trained on, 5 CT-To-MR CGANs were created, where each model was trained on 80\% of the data and produced derived MR slices for the remaining 20\% of the data not seen during model training.

\begin{figure*}[ht]
  \includegraphics[width=1.0\linewidth]{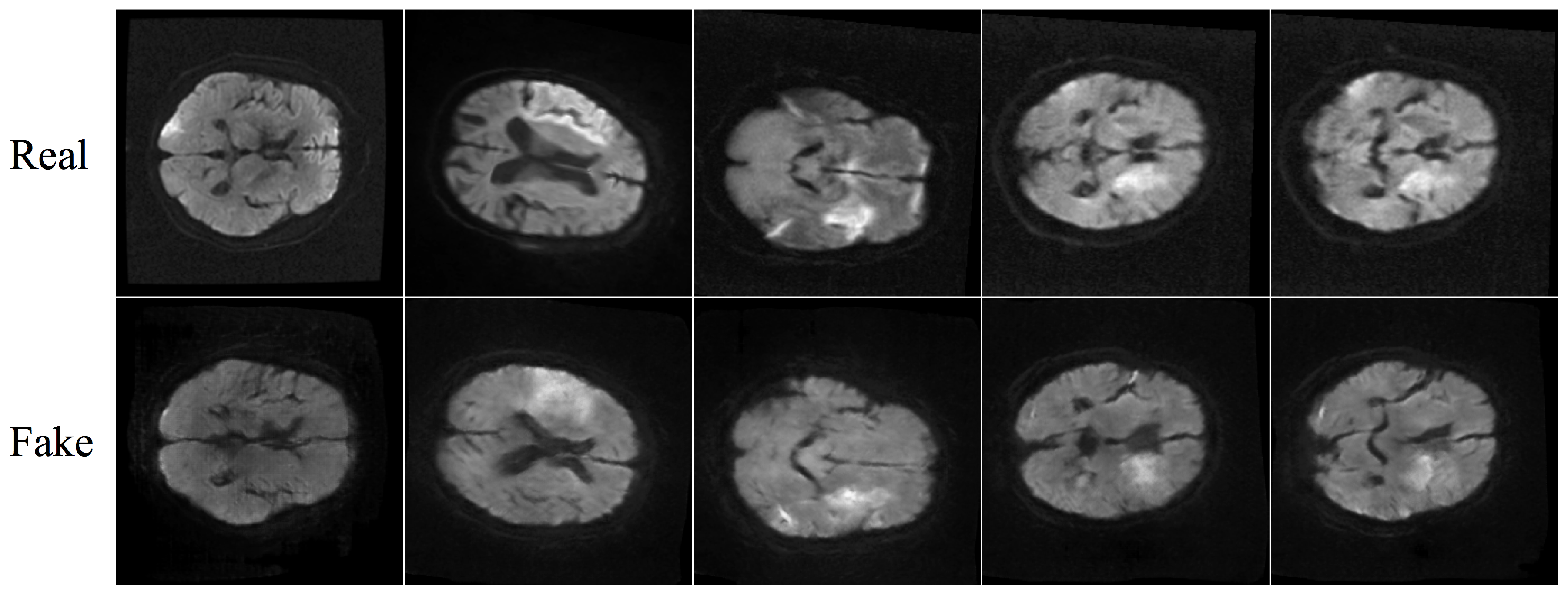}
  \caption{Examples of real and derived MR slices. The top row shows real images and the bottom row shows slices generated from the CT-To-MR Conditional GAN model.}
  \label{fig:brains}  
\end{figure*}

\subsubsection{CGAN Implementation details}

Training of the CT-To-MR CGAN took place by alternating one gradient descent step of the discriminator, followed by one gradient descent step for the generator. A batch size of 1 was used for training all networks. A dropout rate of 0.5 was applied within each ResNet block in the generator (see Fig.~\ref{fig:arch}). Within the final loss function, $G^*$, a value of $\lambda = 100$ was used to weight the combination of both L1 loss and that supplied from $\mathcal{L}_{CGAN}$. Adam optimization used for training both the generator and discriminator with learning rates set to 2$e^{-4}$ and momentum parameters $\beta_1 = 0.5$, $\beta_2 = 0.999$. Affine data transformations consisting of translation, rotation and scaling were used for augmentation. Each network was trained for a total of 200 epochs using PyTorch on a single Nvidia P100 GPU.

\section{Ischemic Core Segmentation FCN Model}
\label{sec:fcn}

The final ischemic core segmentation network is based on the network architecture defined in~\cite{abulnaga2018ischemic}. The model employs a fully convolutional neural network and utilizes pyramid pooling~\cite{zhao2016pyramid} for capturing global and local context. The FCN component of the architecture relies on residual connections~\cite{he2016deep} to aid information flow during training and dilated convolution~\cite{yu2015multi} to cover larger receptive field sizes from the network inputs. Focal loss~\cite{lin2018focal} is used as the loss function to attempt to learn the varying shapes of the lesion masks and effectively deal with the class imbalance between ischemic core and non-infarct areas.

The network is trained using transfer learning, beginning with weights that have been trained on natural images from the Pascal Visual Object Classes dataset~\cite{everingham2010pascal}. During training, data augmentations are created using standard affine transformations including rotation $[-10^\circ, 10^\circ]$, translation $[-10\%, 10\%]$ and scaling $[0.9, 1.1]$.

Two FCNs were trained and compared to perform the final ischemic core segmentation. The network architecture and training details remained the same and the only difference between the networks were the inputs that were fed to them. Inputs to the first network (FCN) consisted of 5-channel 2D slices containing the CT image, together with its corresponding CBF, CBV, TTP and MTT perfusion maps. Inputs to the second network (FCN-GAN) were augmented with an extra channel of information that contained the \emph{derived} MR slice -- generated by the CT-to-MR GAN, conditioned on the 5-channel CTP input.
    
\section{Results}

\begin{figure*}[ht]
  \includegraphics[width=1.0\linewidth]{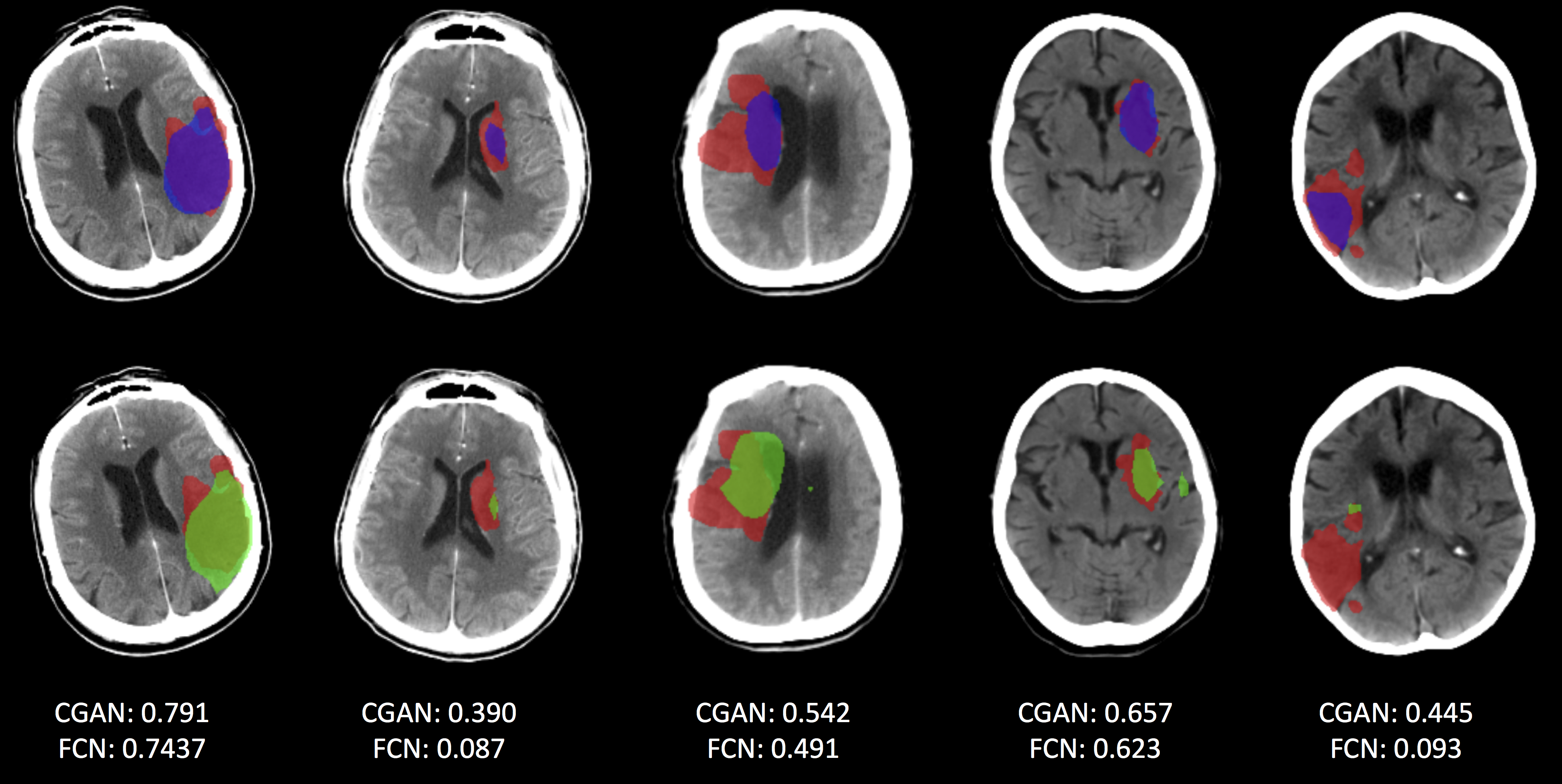}
  \caption{A comparison of segmentation results for scans where the FCN-CGAN model improves upon the FCN model. Ground truth is shown in red. The bottom row shows results produced by the FCN baseline in green and the top row shows results produced by the FCN-CGAN in blue. Also shown are the dice coefficient values for each approach. Note: the dice score is per scan from which each slice was taken from.}
  \label{fig:seg1}
\end{figure*}

\begin{figure*}[ht]
  \includegraphics[width=1.0\linewidth]{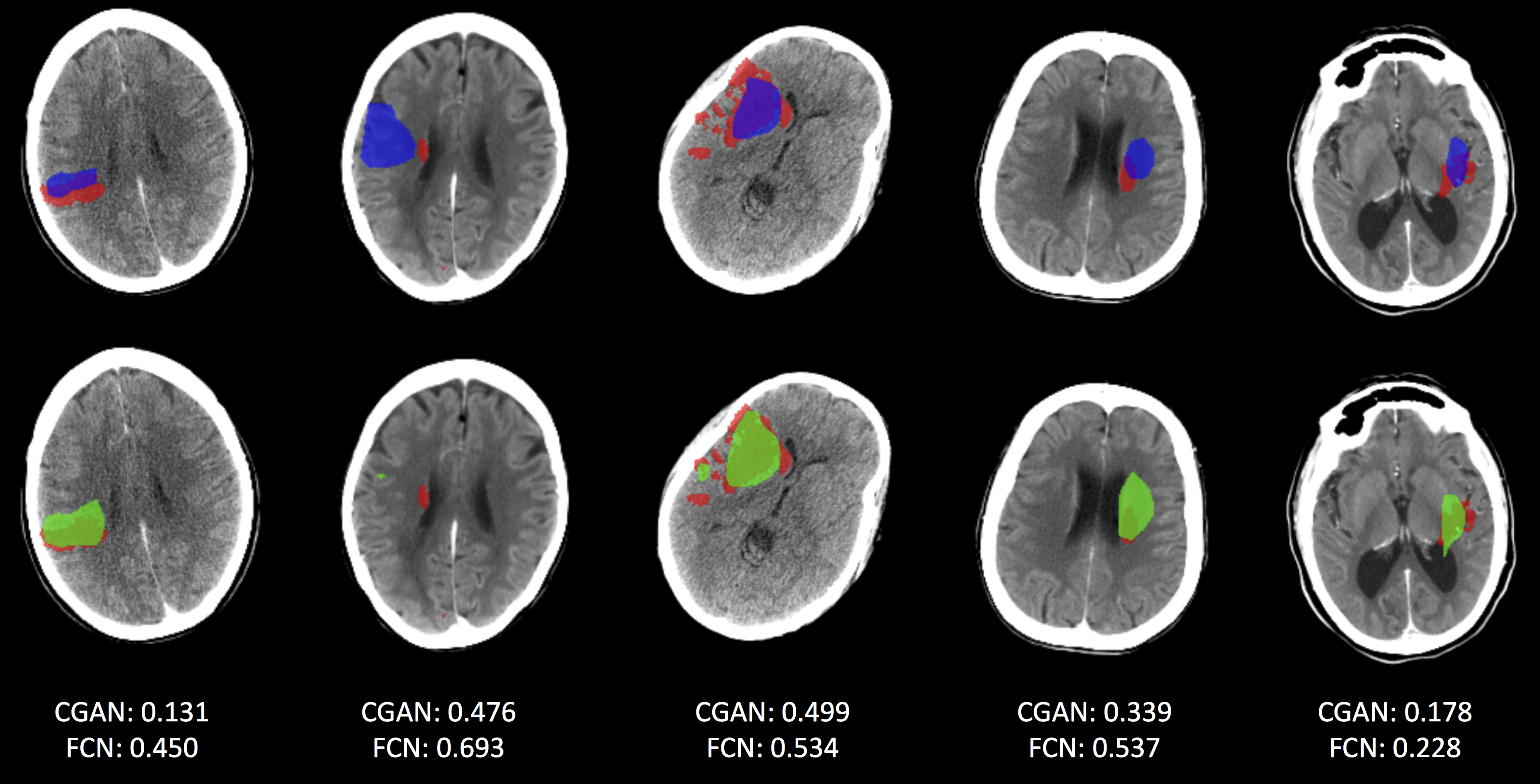}
  \caption{A comparison of segmentation results for scans where the FCN model performs better than the FCN-CGAN model. As in Fig.~\ref{fig:seg1}, the top row shows results from the FCN-CGAN and the bottom row shows those from the FCN model. Coloring information is the same as in Fig.~\ref{fig:seg1}. Overall dice score for each scan is also shown.}
  \label{fig:seg2}
\end{figure*}

We present qualitative results produced by the CT-To-MR CGAN and quantitative results of training FCN models with and without derived MR inputs. Fig.~\ref{fig:brains} shows a subset of results created by the CT-To-MR CGAN model. Five MR slices are shown that were generated by conditioning on CTP input from a 20\% test-set. The top row shows the ground truth MR slice and the bottom row shows the corresponding slice that was created by the CT-To-MR generator. While only a qualitative assessment, it can be seen that hyperintense regions within the ground truth DWI image are approximately replicated by the conditional generator.

For quantitative analysis, Table~\ref{tab:results} compares the 5-fold cross validation results of two ischemic core segmentation models. The first baseline model, FCN, shows the results for a FCN trained only on the 5 original CTP input modalities. FCN-CGAN refers to the same network architecture, but trained by including the generated MR slice as one of the input modalities. Dice coefficient, Hausdorff distance, average distance, precision, recall and absolute volume difference metrics are provided. It can be seen that including derived MR slices from the CT-To-MR CGAN model leads to at least some improvement in overall value for all evaluation metrics.

\begin{table*}
    \caption{Ischemic stroke lesion segmentation results using a FCN baseline model compared to FCN-CGAN that incorporates CT-To-MR input information learned via a conditional adversarial network. Arrows in the columns indicate whether lower or higher values are better. For all evaluation metrics, the FCN-CGAN results in an improved value compared to the FCN baseline.}    
  \centering
  \begin{tabular}{|l|rcl|rcl|}
    \hline
    \textbf{Metric} & \multicolumn{3}{|c|}{\textbf{FCN}}  & \multicolumn{3}{|c|}{\textbf{FCN-CGAN}} \\
    \hline
    $\uparrow$ Dice 						& $0.53$ 				& $\pm$ & 0.25 		& $\mathbf{0.54}$ 	& $\pm$ 	& 0.23 \\
    $\downarrow$ Hausdorff Distance 		& $27.94$				& $\pm$ & 20.27		& $\mathbf{27.88}$ 	& $\pm$ 	& 21.00 \\
    $\downarrow$ Average Distance 			& $4.73$ 				& $\pm$ & 9.88			& $\mathbf{4.37}$ 	& $\pm$  	& 9.35 \\
    $\uparrow$ Precision 					& $0.56$ 				& $\pm$ & 0.27			& $\mathbf{0.56}$ 	& $\pm$  	& 0.25 \\
    $\uparrow$ Recall 					& $0.62$ 				& $\pm$ & 0.27			& $\mathbf{0.63}$ 	& $\pm$  	& 0.25 \\
    $\downarrow$ Absolute Volume Difference 	& $11.53$ 			& $\pm$ & 13.11		& $\mathbf{10.20}$ 	& $\pm$  	& 13.10 \\
    \hline    
  \end{tabular}
  \label{tab:results}
\end{table*}

Fig.~\ref{fig:seg1} shows a sample of segmentation results produced by the FCN and FCN-CGAN approaches. The bottom row shows ischemic core segmentation masks produced by the FCN baseline model in green. Ground truth segmentations are shown in red. The top row compares segmentation results from the FCN-CGAN model in blue to ground truth, once again shown in red. It can be seen that in some cases (e.g.~the 4th case from the right in Fig.~\ref{fig:seg1}) that the FCN-CGAN model was able to correct for islands of false positives that were predicted by the FCN baseline. In addition to the mask predictions, also displayed in Fig.~\ref{fig:seg1} are the overall dice coefficient values for the corresponding scan from which the slice was taken.

Of course, given the overall moderate performance improvement of the FCN-CGAN model, there do also exist scans where the FCN model performs better than the FCN-CGAN model. Fig.~\ref{fig:seg2} shows these cases, where once again the top row represents FCN-CGAN lesion predictions in blue and the bottom row shows FCN lesion predictions in green. We see an interesting failure case in the 2nd scan from the right in Fig.~\ref{fig:seg2}, where both models have failed to predict the ground truth ischemic core region in red, however the FCN-CGAN model generates a much larger over-prediction than the FCN model for this case.

\section{Conclusions \& Future Work}

We have presented an approach that utilized conditional generative adversarial networks to improve the performance of fully convolutional stroke lesion segmentation networks.
Diffusion-weighted magnetic resonance imaging is considered most accurate for early detection of acute stroke \cite{biesbroek2013diagnostic,GILLEBERT2014CTanalysis}, as infarcted brain tissue can be recognized as hyperintense regions of the DWI map compared to surrounding brain tissue. Hence, the motivation for this work was to emulate a DWI map, conditioned on a given CT perfusion input. We aimed to improve the training by generating corresponding DWI maps and including them as inputs to the FCN. Our qualitative results show that CGANs reliably map infarcted core regions to hyperintensities that align with corresponding areas in ground truth MR. We also show quantitative improvements in segmentation performance, as measured by dice coefficient, Hausdorff distance, average distance, precision, recall and absolute volume difference, when including generated MR images as input to the training of segmentation networks. While DWI is currently the gold standard for ischemic stroke lesion segmentation, we demonstrate that CTP images offer detailed information that can be used to generate the necessary hyperintensities for segmenting lesions.

A future direction of this work is to synthesize additional training examples by building generative models~\cite{lau2018scargan,jin2018ct} to generate new examples that can capture the large variability in shape, appearance, and location of pathological tissue. This has the advantage of being able to bolster limited training datasets by synthesizing artificial examples. Many other tasks in medical imaging could similarly benefit from such an approach.

\bibliographystyle{plain}
\bibliography{ichi}

\begin{thebibliography}{10}

\bibitem{abulnaga2018ischemic}
S~Mazdak Abulnaga and Jonathan Rubin.
\newblock Ischemic stroke lesion segmentation in ct perfusion scans using
  pyramid pooling and focal loss.
\newblock {\em arXiv preprint arXiv:1811.01085}, 2018.

\bibitem{biesbroek2013diagnostic}
JM~Biesbroek, JM~Niesten, JW~Dankbaar, GJ~Biessels, BK~Velthuis, JB~Reitsma,
  and IC~Van Der~Schaaf.
\newblock Diagnostic accuracy of ct perfusion imaging for detecting acute
  ischemic stroke: a systematic review and meta-analysis.
\newblock {\em Cerebrovascular diseases}, 35(6):493--501, 2013.

\bibitem{cereda2016benchmarking}
Carlo~W Cereda, S{\o}ren Christensen, Bruce~CV Campbell, Nishant~K Mishra,
  Michael Mlynash, Christopher Levi, Matus Straka, Max Wintermark, Roland
  Bammer, Gregory~W Albers, et~al.
\newblock A benchmarking tool to evaluate computer tomography perfusion infarct
  core predictions against a dwi standard.
\newblock {\em Journal of Cerebral Blood Flow \& Metabolism},
  36(10):1780--1789, 2016.

\bibitem{dai2018scan}
Wei Dai, Nanqing Dong, Zeya Wang, Xiaodan Liang, Hao Zhang, and Eric~P Xing.
\newblock Scan: Structure correcting adversarial network for organ segmentation
  in chest x-rays.
\newblock In {\em Deep Learning in Medical Image Analysis and Multimodal
  Learning for Clinical Decision Support}, pages 263--273. Springer, 2018.

\bibitem{everingham2010pascal}
Mark Everingham, Luc Van~Gool, Christopher~KI Williams, John Winn, and Andrew
  Zisserman.
\newblock The pascal visual object classes (voc) challenge.
\newblock {\em International journal of computer vision}, 88(2):303--338, 2010.

\bibitem{GILLEBERT2014CTanalysis}
C�line~R. Gillebert, Glyn~W. Humphreys, and Dante Mantini.
\newblock Automated delineation of stroke lesions using brain ct images.
\newblock {\em NeuroImage: Clinical}, 4:540 -- 548, 2014.

\bibitem{NIPS2014_5423}
Ian Goodfellow, Jean Pouget-Abadie, Mehdi Mirza, Bing Xu, David Warde-Farley,
  Sherjil Ozair, Aaron Courville, and Yoshua Bengio.
\newblock Generative adversarial nets.
\newblock In Z.~Ghahramani, M.~Welling, C.~Cortes, N.~D. Lawrence, and K.~Q.
  Weinberger, editors, {\em Advances in Neural Information Processing Systems
  27}, pages 2672--2680. Curran Associates, Inc., 2014.

\bibitem{he2016deep}
Kaiming He, Xiangyu Zhang, Shaoqing Ren, and Jian Sun.
\newblock Deep residual learning for image recognition.
\newblock In {\em Proceedings of the IEEE conference on computer vision and
  pattern recognition}, pages 770--778, 2016.

\bibitem{ioffe2015batch}
Sergey Ioffe and Christian Szegedy.
\newblock Batch normalization: Accelerating deep network training by reducing
  internal covariate shift.
\newblock {\em arXiv preprint arXiv:1502.03167}, 2015.

\bibitem{isola2016image}
Phillip Isola, Jun-Yan Zhu, Tinghui Zhou, and Alexei~A Efros.
\newblock Image-to-image translation with conditional adversarial networks.
\newblock {\em arXiv preprint arXiv:1611.07004}, 2016.

\bibitem{jin2018ct}
Dakai Jin, Ziyue Xu, Youbao Tang, Adam~P Harrison, and Daniel~J Mollura.
\newblock Ct-realistic lung nodule simulation from 3d conditional generative
  adversarial networks for robust lung segmentation.
\newblock {\em arXiv preprint arXiv:1806.04051}, 2018.

\bibitem{lau2018scargan}
Felix Lau, Tom Hendriks, Jesse Lieman-Sifry, Sean Sall, and Dan Golden.
\newblock Scargan: chained generative adversarial networks to simulate
  pathological tissue on cardiovascular mr scans.
\newblock In {\em Deep Learning in Medical Image Analysis and Multimodal
  Learning for Clinical Decision Support}, pages 343--350. Springer, 2018.

\bibitem{lin2018focal}
Tsung-Yi Lin, Priyal Goyal, Ross Girshick, Kaiming He, and Piotr Doll{\'a}r.
\newblock Focal loss for dense object detection.
\newblock {\em IEEE transactions on pattern analysis and machine intelligence},
  2018.

\bibitem{louvbld1997ischemic}
Karl-Olof L{\"o}uvbld, Alison~E Baird, Gottfried Schlaug, Andrew Benfield,
  Bettina Siewert, Barbara Voetsch, Ann Connor, Cara Burzynski, Robert~R
  Edelman, and Steven Warach.
\newblock Ischemic lesion volumes in acute stroke by diffusion-weighted
  magnetic resonance imaging correlate with clinical outcome.
\newblock {\em Annals of Neurology: Official Journal of the American
  Neurological Association and the Child Neurology Society}, 42(2):164--170,
  1997.

\bibitem{mahapatra2018deformable}
Dwarikanath Mahapatra, Bhavna Antony, Suman Sedai, and Rahil Garnavi.
\newblock Deformable medical image registration using generative adversarial
  networks.
\newblock In {\em 2018 IEEE 15th International Symposium on Biomedical Imaging
  (ISBI 2018)}, pages 1449--1453. IEEE, 2018.

\bibitem{maier2017isles}
Oskar Maier, Bjoern~H Menze, Janina von~der Gablentz, Levin H{\"a}ni, Mattias~P
  Heinrich, Matthias Liebrand, Stefan Winzeck, Abdul Basit, Paul Bentley, Liang
  Chen, et~al.
\newblock Isles 2015-a public evaluation benchmark for ischemic stroke lesion
  segmentation from multispectral mri.
\newblock {\em Medical image analysis}, 35:250--269, 2017.

\bibitem{mirza2014conditional}
Mehdi Mirza and Simon Osindero.
\newblock Conditional generative adversarial nets.
\newblock {\em arXiv preprint arXiv:1411.1784}, 2014.

\bibitem{nie2017medical}
Dong Nie, Roger Trullo, Jun Lian, Caroline Petitjean, Su~Ruan, Qian Wang, and
  Dinggang Shen.
\newblock Medical image synthesis with context-aware generative adversarial
  networks.
\newblock In {\em International Conference on Medical Image Computing and
  Computer-Assisted Intervention}, pages 417--425. Springer, 2017.

\bibitem{oksuz2018cardiac}
Ilkay Oksuz, James Clough, Aurelien Bustin, Gastao Cruz, Claudia Prieto, Rene
  Botnar, Daniel Rueckert, Julia~A Schnabel, and Andrew~P King.
\newblock Cardiac mr motion artefact correction from k-space using deep
  learning-based reconstruction.
\newblock In {\em International Workshop on Machine Learning for Medical Image
  Reconstruction}, pages 21--29. Springer, 2018.

\bibitem{ulyanovinstance}
D~Ulyanov, A~Vedaldi, and VS~Lempitsky.
\newblock Instance normalization: the missing ingredient for fast stylization.
  corr abs/1607.0 (2016), 2016.

\bibitem{van1994water}
Peter van Gelderen, Marloes~HM de~Vleeschouwer, Daryl DesPres, James Pekar,
  Peter~CM van Zijl, and Chrit~TW Moonen.
\newblock Water diffusion and acute stroke.
\newblock {\em Magnetic Resonance in Medicine}, 31(2):154--163, 1994.

\bibitem{winzeck2018isles}
Stefan Winzeck, Arsany Hakim, Richard McKinley, Jos{\'e}~AADSR Pinto, Victor
  Alves, Carlos Silva, Maxim Pisov, Egor Krivov, Mikhail Belyaev, Miguel
  Monteiro, et~al.
\newblock Isles 2016 and 2017-benchmarking ischemic stroke lesion outcome
  prediction based on multispectral mri.
\newblock {\em Frontiers in neurology}, 9, 2018.

\bibitem{wolterink2017deep}
Jelmer~M Wolterink, Anna~M Dinkla, Mark~HF Savenije, Peter~R Seevinck,
  Cornelis~AT van~den Berg, and Ivana I{\v{s}}gum.
\newblock Deep mr to ct synthesis using unpaired data.
\newblock In {\em International Workshop on Simulation and Synthesis in Medical
  Imaging}, pages 14--23. Springer, 2017.

\bibitem{wolterink2018blood}
Jelmer~M Wolterink, Tim Leiner, and Ivana Isgum.
\newblock Blood vessel geometry synthesis using generative adversarial
  networks.
\newblock {\em arXiv preprint arXiv:1804.04381}, 2018.

\bibitem{yu2015multi}
Fisher Yu and Vladlen Koltun.
\newblock Multi-scale context aggregation by dilated convolutions.
\newblock {\em arXiv preprint arXiv:1511.07122}, 2015.

\bibitem{zhao2016pyramid}
Hengshuang Zhao, Jianping Shi, Xiaojuan Qi, Xiaogang Wang, and Jiaya Jia.
\newblock Pyramid scene parsing network.
\newblock In {\em IEEE Conf. on Computer Vision and Pattern Recognition
  (CVPR)}, pages 2881--2890, 2017.

\bibitem{CycleGAN2017}
Jun-Yan Zhu, Taesung Park, Phillip Isola, and Alexei~A Efros.
\newblock Unpaired image-to-image translation using cycle-consistent
  adversarial networks.
\newblock In {\em Computer Vision (ICCV), 2017 IEEE International Conference
  on}, 2017.

\end{thebibliography}

\end{document}